\documentclass[conference,letter]{IEEEtran}
\usepackage{times,amsmath,epsfig,latexsym,amssymb,psfrag}
\usepackage{amsthm}
\usepackage{cite}
\usepackage{algorithm}
\usepackage{algorithmic}
\usepackage{graphics, color, psfrag,amsfonts}
\usepackage{mathtools}
\usepackage{MnSymbol}
\usepackage{dsfont}
\usepackage{tikz}
\usepackage{amssymb}
\usepackage{pifont}
\theoremstyle{definition}
\newtheorem{definition}{Definition}
 
 \newtheorem{remark}{Remark}

\newtheorem{exmp}{Example}

\begin{document}

\sloppy

\title{\fontsize{22.5}{25}\selectfont Inter-session Network Coding for Transmitting Multiple Layered Streams over Single-hop Wireless Networks}

 \author{
   \IEEEauthorblockN{Mohammad Esmaeilzadeh and Neda Aboutorab}
   \IEEEauthorblockA{Research School of Information Sciences and Engineering, The Australian National University\\Canberra, ACT 0200, Australia\\
Email: \texttt{\{mohammad.esmaeilzadeh,neda.aboutorab\}@anu.edu.au}}
 }



\maketitle

\begin{abstract}
This paper studies the problem of transmitting multiple independent layered video streams over single-hop wireless networks using network coding (NC). We combine feedback-free random linear NC (RLNC) with unequal error protection (UEP) and our goal is to investigate the benefits of coding across streams, i.e. inter session NC. To this end, we present a transmission scheme that in addition to mixing packets of different layers of each stream (intra-session NC), mixes packets of different streams as well. Then, we propose the analytical formulation of the layer decoding probabilities for each user and utilize it to define a theoretical performance metric. Assessing this performance metric under various scenarios, it is observed that inter-session NC improves the trade-off among the performances of users.  Furthermore, the analytical results show that the throughput gain of inter-session NC over intra-session NC increases with the number of independent streams and also by increasing packet error rate, but degrades as network becomes more heterogeneous.
\end{abstract}

\section{Introduction}

The last decade has witnessed great advances of network coding (NC) theory in different areas of wired and wireless communications. Thanks to its capabilities in improving bandwidth utilization and reducing transmission delay and energy, NC has fitted well into numerous applications, from file and media transfer to sensor networks and also distributed storage systems~\cite{Chou:2007:IEEE-SPM-NCI}. However, with the constant demand for better quality of services in such applications and the consequent technology growth, new challenges in NC research are still emerging. 
One area that has been very attractive recently is NC for video streaming~\cite{Magli:2013:IEEE-TMM:NCM}. 

In video streaming, delivering reliable and high quality video is of great interest, but this is often hindered by delay, packet loss and bandwidth limitations. These challenges are even more restrictive when video is transmitted over wireless networks. To deal with these challenges, video streaming standards have been equipped with a number of useful features. For instance, the scalable video coding (SVC) of H.264~\cite{Schwarz:2007:IEEE:SCV} provides layered video streams with various levels of quality, which can be useful when heterogeneity in users' reception capabilities or displays exists. While the added features can alleviate the video streaming challenges to some extent, incorporating NC techniques has shown to provide even more benefits~\cite{Hulya:2009:IEEE-JSAC:VAO, Nguyen:2010:VSN, Nguyen:2011:IEEE-TVT:JNC,Thomos:2011:IEEE-TMM:PDV, Vukobratovic:2012:IEEE-TCOM:UEP}. 

As a case in point, we can refer to~\cite{Hulya:2009:IEEE-JSAC:VAO}, where video-aware opportunistic NC over wireless networks was proposed. In this study, importance of each video packet was first determined based on its deadline and contribution to overall video quality. Then, considering the decodability of packets by several users, efficient network codes were selected to maximize the overall video quality. It was shown that the proposed scheme significantly outperforms scheduling algorithms without NC. Another research that similarly considered quality and deadline of video packets was conducted in~\cite{Nguyen:2011:IEEE-TVT:JNC}. In this study, the authors proposed to use the finite horizon Markov decision process (MDP) to select efficient network codes, not only by considering the next transmission, but also by taking into account all the transmissions before packets' deadline. Their scheme showed to provide even extra gain over non-NC schemes in their multiuser single-hop wireless network.

In~\cite{Hulya:2009:IEEE-JSAC:VAO} and~\cite{Nguyen:2011:IEEE-TVT:JNC}, which are discussed above, XOR-based NC was used. While this type of NC has many favorable characteristics, the dependency of codes selection on packet delivery acknowledgments (feedback) makes it unsuitable for some systems/networks. Hence, another type of NC, namely random linear NC (RLNC) with less dependency on feedback, has been studied for video streaming as well~\cite{Nguyen:2010:VSN,Thomos:2011:IEEE-TMM:PDV, Vukobratovic:2012:IEEE-TCOM:UEP}. In these studies, the authors have utilized the SVC and have proposed to combine unequal error protection (UEP)~\cite{Ha:2008:LUE} with RLNC (UEP+RLNC) to achieve improved performance over non-NC schemes. Moreover,~\cite{Nguyen:2010:VSN, Vukobratovic:2012:IEEE-TCOM:UEP} proposed more analytical approaches and obtained the decoding probabilities of different layers of SVC. They also showed that inter-layer UEP+RLNC outperforms intra-layer UEP+RLNC in their considered single stream system setup.

In this paper, we take one step forward and consider transmission of multiple layered video streams over single-hop wireless networks. We use UEP+RLNC and aim to investigate the gain of coding across streams, which we refer to as \emph{inter-session NC}. To this end, we present a transmission scheme that mixes not only packets of different layers of each stream (intra-session NC), but also packets of different streams (inter-session NC) to benefit from overhearing. Then, we propose an analytical approach to calculate the decoding probabilities of different layers for every user. This is the main contribution of this paper. Finally, the best transmission scheme that gives the optimum overall performance is obtained.

It is worth mentioning that a few studies with generally similar focus on inter-session network coding are available in the literature~\cite{Tran:2009:IEEE-JSAC:HNC, Hulya:2011:INFOCOM:I2NC,Bourtsoulatze:2013:ONLINE}. However, those studies differ from this work from various aspects. Firstly, these works have used XOR-based NC. Secondly, the authors in~\cite{Bourtsoulatze:2013:ONLINE} have focused on wireline mesh networks. Moreover, the studies in~\cite{Hulya:2011:INFOCOM:I2NC, Tran:2009:IEEE-JSAC:HNC} have not used \emph{layered} data and finally, the study in~\cite{Tran:2009:IEEE-JSAC:HNC} has not considered any deadline constraint for delivering the packets, which we shall consider for our video streaming application. Hence, the problem we are studying in this paper is novel and has not been addressed previously.  

The remainder of the paper is organized as follows. The system model is presented in Section II. In Section III, we formulate the decoding probabilities of different layers and define the performance metric. Section IV provides the numerical results, and finally Section V concludes the paper.

\section{System Model}

The system model consists of a sender and $N$ wireless users. The users are assumed to be heterogeneous, which means the channels between the users and the sender are not identical and have packet error rates (PERs) of $P_{e_i},~1\leq i\leq N$. The sender is supposed to transmit to each user a unique layered stream, i.e. to deliver a layered video stream $i$ to the $i-$th user. 

These video streams are considered to be chunked, where each chunk corresponds to a fixed number of frames that we refer to as a group of picture (GOP), to be compliant with video streaming standards. It is assumed that GOPs of different streams have equal duration of $t$ seconds and are synchronized.
Hence, the sender has $t$ seconds in total to deliver one GOP of each stream to its intended user. Therefore, if the sender has transmission rate of $r$ packets per second, the possible total number of packet broadcast for the $N$ streams per GOP is $N_t=tr$. These $N_t$ transmissions, which are considered to be fixed and limited, are dedicated to network coded packets and are explained in the next subsections.

We assume a layered video stream $i$ has $L_i$ layers and use $K^i=[k_1^i,k_2^i,...,k_{L_i}^i]$ to represent the number of packets in different layers per GOP, where $k_{\ell}^i$ denotes the number of packets for the $\ell-$th layer ($1\leq \ell \leq L_i$). Note that based on the video content of stream $i$ at different times, these $k_\ell^i, 1\leq \ell \leq L_i$ can take different values for different GOPs. For all the streams, the first layer is considered to be the base layer and consequently the most important one and the last layer is the one with the least importance. 

\subsection{Random Linear Network Coding (RLNC)} \label{RLNC}

\begin{figure}[!t]
\centering
\includegraphics[width=1.95in]{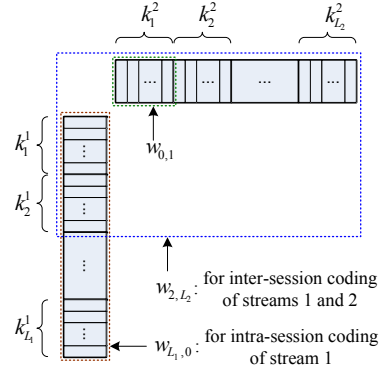}
\vspace{-2mm}
\caption{Example of intra- and inter-session coding windows for 2 independent layered streams. } \label{Fig1}
\end{figure}

We utilize RLNC~\cite{Ho:IEEE-IT:2006:RLN} in this study and in addition to inter-layer coding for each stream (i.e. intra-session NC), we benefit from coding across streams (i.e. inter-session NC). In this approach, $W_{\text{tot}}=\prod_{i=1}^N(L_i+1)-1$ windows for coding are considered, where window $w_{\ell_1,\ell_2,...,\ell_N}$ ($0\leq \ell_i \leq L_i$) contains all the packets up to layer $\ell_1$ from stream $1$ and up to layer $\ell_2$ from stream $2$ and so on. Then, based on the transmission policy, network coded packets from different windows are generated and transmitted, which results in UEP+RLNC. It can be inferred that an all-zero index window is undefined. Moreover, a window with only one nonzero index, e.g. only the $i-$th index is nonzero, corresponds to intra-session NC of only stream $i$, and windows with at least two nonzero indices correspond to the inter-session NC. This approach can be considered as an $N$ dimensional extension to the expanding windows (EW) UEP coding presented in~\cite{Vukobratovic:2012:IEEE-TCOM:UEP}. A 2-D example of the coding windows is depicted in Fig.~\ref{Fig1}.

The theory behind RLNC encoding and decoding has been studied comprehensively in the literature during the past decade, with the effect of field size also discussed quite in details (e.g.~\cite{Lucani:2009:GLOBECOM:RNC,Esmaeilzadeh:2014:IEEE-TCOM:JOP}). Hence, we are not going to elaborate these here. We assume that the coding coefficients are nonzero and are chosen randomly from large enough field sizes, which results in the following remarks:

\begin{remark} \label{remark1}
The coded packets generated from packets of a coding window are all linearly independent (with high probability, which we approximate to be one).
\end{remark}
\begin{definition}
Considering two coding windows $w_a$ and $w_b$ with $x_a$ and $x_b$ packets, respectively, we define $w_a$ to be a subset of $w_b$ and denote it by $w_a\subset w_b$, if $w_b$ contains all the $x_a$ packets of $w_a$.
\end{definition}
According to this definition, for the example in Fig.~\ref{Fig1}, $w_{0,1}\subset w_{2,L_2}$.
\begin{remark} \label{remark2}
For two coding windows $w_a$ and $w_b$, if $w_a\subset w_b$, then a coded packet generated from $w_a$ is linearly independent from all the coded packets generated from $w_b$. The reason behind this is that all the coding coefficients are considered to be nonzero.
\end{remark}
\begin{remark} \label{remark3}
In order to decode all the $x_b$ packets of a coding window $w_b$, $x_b$ linearly independent coded packets from $w_b$ are required. If $w_a\subset w_b$, according to Remark~\ref{remark2}, coded packets from $w_a$ can also be used for decoding, but only to a maximum of $x_a$ coded packets.
\end{remark}

These remarks are used to obtain the decoding probabilities for our approach in the next section. It is worth noting that one can incorporate the effect of filed size in decoding probabilities in our method using a similar approach as in~\cite{Esmaeilzadeh:2014:IEEE-TCOM:JOP}.

\subsection{Transmission Schemes}

As mentioned previously, based on the system parameters, we assume the sender (for the duration of one GOP and for all the $N$ streams in total) is allowed to transmit $N_t$ packets. Here we explain how these transmissions are carried out. 

Similar to\cite{Vukobratovic:2012:IEEE-TCOM:UEP, Esmaeilzadeh:2014:IEEE-TCOM:JOP, Esmaeilzadeh:2013:IEEE-PIMRC:QoS}, we assume feedback-free transmissions. Hence, the sender decides in advance on how many coded packets from each window it should transmit, and then sends them one after another, without waiting for any feedback. Assuming that $n_{\ell_1,\ell_2,...,\ell_N}^t$ RLNC packets are generated from the packets in the window $w_{\ell_1,\ell_2,...,\ell_N}$, then $\sum_{\ell_1=0}^{L_1}\sum_{\ell_2=0}^{L_2}\cdots\sum_{\ell_N=0}^{L_N}n_{\ell_1,\ell_2,...,\ell_N}^t=N_t$. We call $T=[n_{\ell_1,\ell_2,...,\ell_N}^t]$ a feedback-free transmission policy with inter-session coding. It is evident that if RLNC packets are generated only from windows corresponding to intra-session coding, as explained in Section~\ref{RLNC}, then $T$ is a feedback-free transmission policy with intra-session coding. The decision on the optimum policies is made based on an aggregate function of users' performance by taking into account the number of transmissions $N_t$ and long-term channel characteristics. This will be briefly discussed later in Section~\ref{section3}. 
 
\subsection{Performance Metric}

In this paper, we consider a theoretical performance metric that is a weighted sum of the probabilities of layer decoding. 
Here, the weights can be selected such that the performance metric reveals the expected throughput or the percentage of the frames decoded,
as will be explained in Section~\ref{section3}. We denote the performance metric for user $i$ by $\eta_i$.

\section{Formulation of Theoretical Performance Metric} \label{section3}

In this section, we will study the formulation of our theoretical performance metric.

Under the assumptions made in previous sections, the main purpose is to formulate and obtain the probability that user $i$ can decode the packets of layer $\ell$ (and of course all the packets of lower layers) of its intended stream. 
We denote these probabilities by $P_{\ell}^i(K^1,K^2,\cdots,K^N,T)$, where $1\leq i\leq N,~0\leq \ell \leq L_i$. Then, we use the weighted sum of these probabilities and define the theoretical performance metric. 
%

To obtain these probabilities, we consider that out of the $n_{\ell_1,\ell_2,...,\ell_N}^t$ transmitted coded packets of the window $w_{\ell_1,\ell_2,...,\ell_N}$, user $i$ has received $n_{\ell_1,\ell_2,...,\ell_N}^r$ packets, where $0\leq n_{\ell_1,\ell_2,...,\ell_N}^r\leq n_{\ell_1,\ell_2,...,\ell_N}^t$. Thus, we denote by $R=[n_{\ell_1,\ell_2,...,\ell_N}^r]$ the number of received packets from different windows. Then, $P_{\ell}^i(K^1,K^2,\cdots,K^N,T)$ can be written as

\begin{align} \label{P_layer}
&P_{\ell}^i(K^1,K^2,\cdots,K^N,T) = \nonumber\\
 &\sum_{\text{all possible }R}P(R|T)I(D_i(K^1,K^2,\cdots,K^N,R)=\ell)
\end{align}
where
\begin{align} \label{Eq2}
P(R|T) =& \prod_{0 \le \ell_1 \le L_1}\prod_{0 \le \ell_2 \le L_2}\cdots \prod_{0 \le \ell_N \le L_N} {{n_{\ell_1,\ell_2,...,\ell_N}^t}\choose{n_{\ell_1,\ell_2,...,\ell_N}^r}}\times \nonumber\\ 
&(1-P_{e_i})^{n_{\ell_1,\ell_2,...,\ell_N}^r} P_{e_i}^{n_{\ell_1,\ell_2,...,\ell_N}^t-n_{\ell_1,\ell_2,...,\ell_N}^r}
\end{align}
is the probability of receiving $R$ when $T$ is transmitted over a channel with PER of $P_{e_i}$.
$I(\cdot)$ is an indicator function with output $1$ if its argument, which is a logical expression, is true. 

The function $D_i(K^1,K^2,\cdots,K^N,R)$ calculates the highest decodable layer for the $i-$th user based on the number of data packets ($K^1,K^2,\cdots\, K^N$) and the number of received packets ($R$). We have proposed Algorithm 1 to calculate the value of this function, based on Remarks~\ref{remark1} to~\ref{remark3}.\footnote{The notation $A(0:b_1,0:b_2,...,0:b_n)$ used in Algorithm 1 represents all the elements of matrix $A$ (or vector $A$ when $n=1$) with the index of the first dimension between $0$ and $b_1$, the index of the second dimension between $0$ and $b_2$ and so on.}

\begin{algorithm} [!t] \label{alg1} 
\caption{Calculate $d_i=D_i(K^1,K^2,\cdots,K^N,R)$}
\begin{algorithmic} 
\STATE 1:~  $d_i \leftarrow 0$;
\STATE 2:~ {\bf for} $j \gets 0:N$ {\bf do}
\STATE 3:~  ~~$K^j \leftarrow [0~~K^j]$; 
\STATE 4:~  {\bf end for}
\STATE 5:~ $True \gets 1$;
\STATE 6:~  {\bf while} $True = 1$ {\bf do}
\STATE 7:~  ~~$True \leftarrow 0$;
\STATE 8:~ ~~{\bf for} $\ell_1 \gets 0:L_1$ {\bf do}
\STATE ~~~~~~~$\ddots$
\STATE 9:~ ~~~~{\bf for} $\ell_N \gets 0:L_N$ {\bf do}
\STATE 10:~ ~~~~~{\bf if} $R(\ell_1,\ell_2,\cdots,\ell_N) > 0$ {\bf then}
\STATE 11: ~~~~~~~~{\bf if} $\sum R(0:\ell_1,0:\ell_2,\cdots,0:\ell_N) \geq$
\STATE \hspace{16.5mm}~ $\sum(K^1(0:\ell_1))+...+\sum(K^N(0:\ell_N))$ {\bf then}
\STATE 12: ~~~~~~~~~~~$d_i\leftarrow\max\{d_i,\ell_i\}$;
\STATE 13: ~~~~~~~~~~~{\bf if} {$d_i=L_i$} {\bf then}
\STATE 14: ~~~~~~~~~~~~~{\bf goto} line $27$;  
\STATE 15: ~~~~~~~~~~~{\bf end if}
\STATE 16: ~~~~~~~~~~~$R(0:\ell_1,0:\ell_2,\cdots,0:\ell_N)\leftarrow0$;
\STATE 17: ~~~~~~~~~~~{\bf for} $j \gets 0:N$ {\do}
\STATE 18: ~~~~~~~~~~~~~$K^j(0:\ell_j) \leftarrow 0$;
\STATE 19: ~~~~~~~~~~~{\bf end for}
\STATE 20: ~~~~~~~~~~~$True \leftarrow 1$;
\STATE 21: ~~~~~~~~~~~{\bf goto} line $6$;  
\STATE 22: ~~~~~~~~{\bf end if}
\STATE 23: \hspace{7.35mm}{\bf end if}
\STATE 24: \hspace{4.33mm}{\bf end for}
\STATE ~~~~~~~$\udots$
\STATE 25: \hspace{2mm}{\bf end for}
\STATE 26: {\bf end while}
\STATE 27: {\bf return} $d_i$;
\end{algorithmic}
\end{algorithm}

The algorithm, at every iteration of the `while' loop, checks the decoding condition (line 11) for different coding windows sequentially (i.e. using the `for' loops). Whenever the condition is met, the function's output $d_i$ is updated (line 12), and if $d_i$ is not yet equal to $L_i$, the elements of $K^1, K^2,\cdots, K^N$ and $R$ corresponding to the current coding window are all set to zero (lines 16-19) and the iteration starts from the beginning. 
The reason behind setting those elements of $K^1, K^2,\cdots, K^N$ and $R$ to zero is to eliminate their effect on consequent iterations and is necessary because without this, in contrast to Remark~\ref{remark3}, cases where a coding window $w_a\subset w_b$ with $x_a$ packets contributes more than $x_a$ packets for decoding of $w_b$ could be possible. This will become more clear by considering Example 1. If in an iteration of the `while' loop the condition is not met for any of the coding windows, the algorithm finishes by reporting the latest $d_i$. 

To make the algorithm more clear, let us consider the following example.
    
\begin{figure}[!t]
\centering
\includegraphics[width=2.83in]{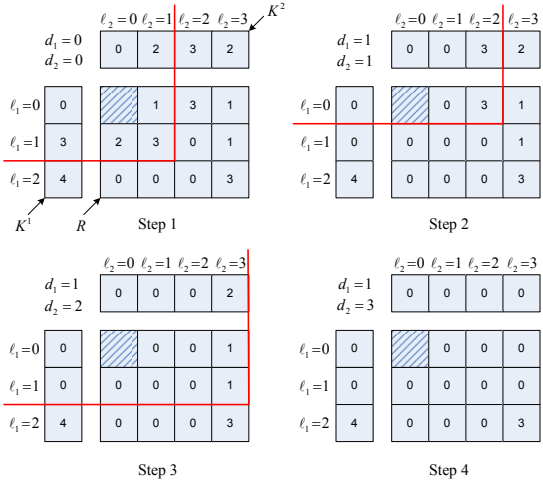}
\vspace{-2mm}
\caption{An example showing how Algorithm 1 obtains the highest decodable layer. The provided $d_1$ and $d_2$ values are prior to running each step.} \label{Fig2}
\end{figure}

\begin{exmp}
Consider Fig.~\ref{Fig2} where we have set $N=2$, $L_1=2$, $L_2=3$, $K^1=[3, 4]$ and $K^2=[2, 3, 2]$. A user (can be either user 1 or 2) has received different number of coded packets from different windows, which are shown in matrix $R$. The goal is to obtain $d_1$ (or $d_2)$. 

First, an element of zero (`0') is attached to the beginning of $K^1$ and $K^2$ (lines 2-4 of Algorithm 1). This zero is inserted to account for cases when a selected coding window is not containing any packet from the 1st or 2nd stream. Having a closer look at Step 1 in Fig.~\ref{Fig2}, which corresponds to the first iteration of the `while' loop in Algorithm 1, it can be observed that the decoding condition is not met for any of $w_{0,1}, w_{0,2}, w_{0,3}$ and $w_{1,0}$, until it is met for $w_{1,1}$ (shown with red lines). Hence, $d_1$ (or $d_2$) is updated, the corresponding elements of $K^1, K^2$ and $R$ (i.e. the elements inside the red lines) are set to zero and the algorithm continues with the next iteration. The iterations continue until $d_i$ reaches $L_i$ or till the decoding condition is not holding any more for any decoding windows, which in this example happens for the 2nd user in Step 3 and for the 1st user in Step 4, respectively.

It is worth noting that if the effect of the already decoded windows has not been eliminated in Steps 1, 2 and 3, the function would have mistakenly returned $d_1=2$ in Step 4, because the decoding condition would have met for $w_{2,3}$. 
\qed
\end{exmp}
 
Having calculated the highest decodable layer of user $i$ using Algorithm 1, the layer decoding probabilities in~\eqref{P_layer} can be obtained. Now, we define the theoretical performance metric as
\begin{align}  \label{Eq3}
\eta_i = \sum_{\ell=1}^{L_i}a_{\ell}P_{\ell}^i(K^1,K^2,\cdots,K^N,T) 
\end{align}
where $a_{\ell}$ reflects the cumulative importance of layers $1$ to $\ell$. For instance, considering the temporal scalability in SVC of H.264, for an $L_i=2$ layers case, if number of frames per layer are equal, with $a_1=0.5$ and $a_2=1$, $\eta_i$ will give the expected percentage of frames decoded. In this paper we only focus on throughput, thus we use $a_\ell=(\sum_{j=1}^\ell k_j^i)/(\sum_{j=1}^{L_i}k_j^i)$ that results in the expected throughput. 

To select a transmission policy, as mentioned previously, we consider an aggregate function of users' performance and maximize it to obtain the optimum policy. While different aggregate functions are possible~\cite{Esmaeilzadeh:2014:IEEE-TCOM:JOP, Esmaeilzadeh:2013:IEEE-PIMRC:QoS}, we use the arithmetic mean $E\{\eta\}=\sum_{i=1}^N\eta_i/ N$. Hence, the optimum transmission policy can be derived as 
\begin{align} \label{T_EQ}
T^*=\arg\max_T\{E\{\eta\}\}
\end{align}
which we obtain by exhaustively searching through all possible cases.

\section{Numerical Results} \label{section4}

%

In this section, we present the numerical results comparing the throughput performance of the inter-session NC with that of the intra-session NC. To calculate the performance of the intra-session NC, we use~\eqref{P_layer} to~\eqref{Eq3} with proper selection of coding windows as described in Section~\ref{RLNC}. We also compare some of the results with an \emph{uncoded} UEP scheme, where $N_t$ transmissions are unevenly dedicated to transmission of original packets from different layers and different streams.

\begin{figure}
\centering
\includegraphics[width=2.85in]{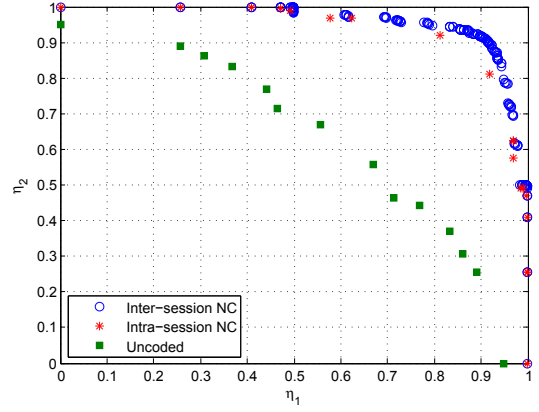}
\vspace{-3mm}
\caption{Performance trade-off for inter- and intra-session NC and uncoded UEP scheme for 2 independent layered streams.} \label{Fig3}
\end{figure}

We start with $N=2$ streams and consider both streams to have 2 layers. We consider a GOP of stream 1 with $K^1=[3, 3]$ and a GOP of stream 2 with $K^2=[3, 3]$, and assume $N_t=17$. Then, obtain $\eta_1$ and $\eta_2$ for all the possible transmission policies (solutions) under $P_{e_1}=P_{e_2}=0.2$. The \emph{Pareto optimal} solutions\footnote{A solution/point $(\eta_1^*,\eta_2^*)$ is called Pareto optimal if no other solution/point $(\eta_1,\eta_2)$ with both $\eta_1>\eta_1^*$ and $\eta_2>\eta_2^*$  exists.} are shown in Fig.~\ref{Fig3}. Since the total number of transmissions is limited, there exist a trade-off between $\eta_1$ and $\eta_2$ for different policies.

It can be observed that both intra- and inter-session NC schemes outperform the uncoded scheme in terms of throughput. Moreover, the inter-session NC has slightly better performance, with more Pareto optimal points. This means that to select a transmission policy, inter-session NC offers more coding options to satisfy both users concurrently.

\begin{figure}[!t]
\centering
\includegraphics[width=2.85in]{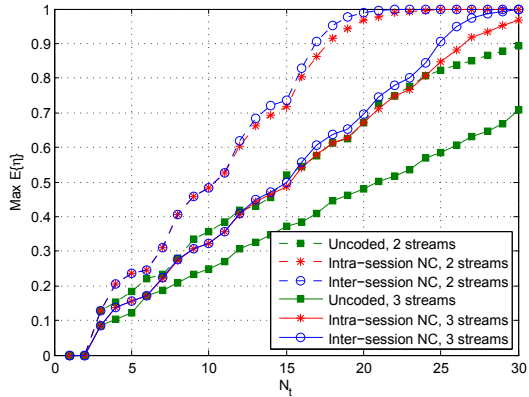}
\vspace{-3mm}
\caption{Performance comparison of inter- and intra-session NC and uncoded UEP scheme for 2 and 3 streams cases.} \label{Fig4}
\end{figure}

Next, we present the maximum $E\{\eta\}$ results by using~\eqref{T_EQ} for $N=2$ and $N=3$ streams cases. In each of these cases, we consider one of the streams to be single layered. Thus, for $N=2$ case, $K^1=[3, 3]$ and $K^2=[6]$ and for $N=3$ case, $K^1=K^2=[3, 3]$ and $K^3=[6]$ are used. Fig.~\ref{Fig4} depicts the results with equal PER of 0.2 considered for all users.  

In $N=3$ case, there are $W_{tot}=17$ coding windows for inter-session NC, from which 5 coding windows are for intra-session NC. To find the optimal inter-session NC, $E\{\eta\}$ for all distributions of $N_t$ transmissions among these 17 windows should be calculated, which can be computationally expensive, especially for large $N_t$ values. Therefore, we decided to utilize a subset of coding windows for inter-session NC, i.e. we used $w_{1,1,0}, w_{1,0,1}, w_{0,1,1}, w_{1,1,1}, w_{2,2,0}, w_{2,2,1}$ in addition to the 5 coding windows of intra-session coding. Although the depicted result in this case is not for the best inter-session NC possible, it still improves the performance of intra-session NC.

Results in Fig.~\ref{Fig4} reveal that inter-session NC improves the throughput performance of intra-session NC over a range of $N_t$ values, and outside this range, they work similarly. This range is affected by PERs, number of packets and also number of streams. Furthermore, it is observed that the more the number of streams, the higher the improvement of inter-session NC over intra-session NC.

So far, we considered PER of different users to be equal. Now, we discuss the effect of unequal PERs. We consider $N=3$ and $K^1=[2, 3], K^2=[3, 3], K^3=[4]$ and obtain the maximum improvement of inter-session NC over intra-session NC for different PERs combinations. Results are provided in Table~\ref{improvement_table}. Note that inter-session NC windows similar to those for $N=3$ case in Fig.~\ref{Fig4} are utilized here as well. It is observed that the inter-session NC offers higher improvements for higher PER values, but the throughput gain diminishes as the heterogeneity among users increases.

\begin{table} [t]
\renewcommand{\arraystretch}{1.2}
\caption{Percentage of maximum improvement of inter-session NC over intra-session NC for different combinations of PERs.}
\label{improvement_table}
\centering
\addtolength{\tabcolsep}{-2pt}
\begin{tabular}{c|ccc|cccc}
		 $P_{e_1}$&$0.1$&$0.2$&$0.3$&$0.3$&$0.2$&$0.2$&$0.1$ \\
		$P_{e_2}$&$0.1$&$0.2$&$0.3$&$0.2$&$0.3$&$0.3$&$0.2$ \\
		 $P_{e_3}$&$0.1$&$0.2$&$0.3$&$0.2$&$0.3$&$0.1$&$0.3$ \\

\hline
Max improvement (\%)& 4.8\% & 6.3\% & 7.5\% & 5.7\% & 5.8\%& 4.2\%& 3.2\%    \\

\end{tabular}
\vspace{ -1.2 mm}
\end{table}

\section{Conclusion}

In this paper, we proposed a novel analytical approach to study the inter-session random linear network coding for transmission of multiple layered streams over wireless networks. We investigated the gain of coding across streams over coding within streams and highlighted the effect of number of transmissions, number of streams, PER and network heterogeneity on this gain. As a part of our ongoing research, we intend to test the proposed approach with real video streams next, which requires more informed decisions about coding windows to handle the existent computational complexities.

\section*{Acknowledgment}

The authors would like to thank Dr. Parastoo Sadeghi for the useful technical discussions and the valuable comments.

This work was supported under the Australian Research Council Discovery Projects and Linkage Projects funding schemes (project nos. DP120100160 and LP100100588).

\ifCLASSOPTIONcaptionsoff
  \newpage
\fi

\bibliographystyle{IEEEtran}
\bibliography{IEEEabrv,Ref}

\end{document}